\documentclass[pdftex,twocolumn,epjc3]{svjour3}

\usepackage{amssymb}
\usepackage{bbm}
\usepackage{hyperref}
\usepackage[normalem]{ulem}

\usepackage[T1]{fontenc}

	\renewcommand{\d}{\mathrm{d}}
	
    \newcommand{\eqref}[1]{(\ref{#1})}

	\newcommand{\msout}[1]{\ifmmode\text{\sout{\ensuremath{#1}}}\else\sout{#1}\fi}

\begin{document}

\title{Rainbows without unicorns: Metric structures in theories with Modified Dispersion Relations}	
	
\newcommand{\addressRoma}{Dipartimento di Fisica, Universit\'{a} ``La Sapienza''
and Sez. Roma1 INFN, P.le A. Moro 2, 00185 Roma, Italy}
\newcommand{\addressICRA}{ICRANet, Piazza della Repubblica 10, I-65122 Pescara, Italy}
\newcommand{\addressCAPES}{CAPES Foundation, Ministry of Education of Brazil, Bras\'ilia, Brazil}
\newcommand{\addressUFPB}{Departamento de F\'{i}sica, Universidade Federal da Para\'{i}ba, Caixa Postal 5008, CEP 58051-970, Jo\~{a}o Pessoa, PB, Brazil}
\newcommand{\addressZagreb}{Ru\dj er Bo\v{s}kovi\'{c} Institute, Division of Theoretical Physics, Bijeni\v{c}ka c.54, HR-10002 Zagreb, Croatia}
\newcommand{\addressMexico}{Instituto de Ciencias Nucleares, Universidad Nacional Aut\'onoma de M\'exico, AP 70543, M\'exico, CDMX 04510, Mexico}

\author{Iarley P. Lobo\thanksref{e1,addr1,addr2,addr3,addr4}, Niccol\'{o} Loret\thanksref{e2,addr5}, Francisco Nettel\thanksref{e3,addr1,addr6}}

\thankstext{e1}{e-mail: iarley\_lobo@fisica.ufpb.br}
\thankstext{e2}{e-mail: niccolo@accatagliato.org}
\thankstext{e3}{e-mail: francisco.nettel@roma1.infn.it}

\institute{ \addressRoma\label{addr1}  \and
\addressICRA\label{addr2} \and
\addressCAPES\label{addr3} \and
\addressUFPB\label{addr4} \and
\addressZagreb\label{addr5} \and
\addressMexico\label{addr6}
}

\maketitle

\begin{abstract}
Rainbow metrics are a widely used  approach to  metric formalism for theories with Modified Dispersion Relations. They have had a huge success in the Quantum Gravity Phenomenology literature, since they allow to introduce momentum-dependent spacetime metrics into the description of systems with Modified Dispersion Relation. In this paper, we introduce the reader to some realizations of this general idea: the original Rainbow metrics proposal, the momentum-space-inspired metric and a Finsler geometry approach. As the main result of this work we also present an alternative definition of a four-velocity dependent metric which allows to handle the massless limit. This paper aims to highlight some of their properties and how to properly describe their relativistic realizations.
%This approach, however, presents some compatibility issues with a relativistic description, even in the case of introducing deformed spacetime symmetries to keep the theory's modified dispersion relation invariant. In this paper we would like to introduce the readers to this issue and to describe how the relativistic properties of the theory can be recovered taking into account a more complex (but also complete) momentum-space curvature framework. We also introduce a new metric structure from a Polyakov-like description of the action.
\end{abstract}

\section{Introduction}  
\label{sec:intro}

The analysis of Planck-scale modified dispersion relations (MDRs) inspired by different approaches to quantum gravity has attracted a lot of attention in recent years \cite{AmelinoCamelia:2008qg,Mattingly:2005re}. The motivation for this comes mainly from the fact that predictions arising from such modifications could be confronted with astrophysical and cosmological observations allowing to test some general features about the quantum nature of spacetime. For instance, we can find that different observations, confronting the detection time of particles with different energy (see for instance \cite{neutrini} and references therein), could be set up in order to put constraints on the deformation parameters characterizing the MDR \cite{Rosati:2015pga,Jacob:2008bw}.\footnote{Other phenomena like reactions threshold violations which may be suggested by a MDR may lead to different predictions in two different scenarios.}

The different predictions for this kind of phenomena can be accommodated in two different scenarios.  On the one hand, there is the Lorentz Invariance Violation (LIV) framework which presupposes an observer-dependent scenario \cite{Jacob:2008bw,Stecker:2014oxa,synchrotron}. On the other hand, a relativistic description for the time delay predictions \cite{AmelinoCamelia:2012it,Rosati:2015pga}, i.e., observer independent, is possible within the Double Special Relativity (DSR) framework introduced in \cite{AmelinoCamelia:2000mn} (see also \cite{KowalskiGlikman:2001gp,Magueijo:2002am}). 

Being cosmology and astrophysics the most suitable arenas to test this kind of theories, it is of paramount importance to take into account the interplay between such deformation effects and spacetime curvature either in the case of the Poincar\'{e} symmetry breakdown or the deformation scenario. Therefore, efforts have been devoted to try to find a geometric characterization of the MDRs. A first attempt to incorporate MDRs into a metric formalism was the so called {\it Rainbow metrics} approach \cite{Rainbow}. In this framework, spacetime metric should be modified according to the particles' modified dispersion relation expressed as $m^2=g^{\mu\nu}(p)p_\mu p_\nu$, leading to a family of energy-dependent metrics $ds^2=g_{\alpha\beta}(p) \d x^\alpha \d x^\beta$ (see for instance \cite{Rainbow,Kimberly:2003hp}), and recently has attracted a lot of interest in the literature (see for instance \cite{Ling:2006az,Ali:2014cpa,Santos:2015sva,Carvalho:2015omv,Garattini:2015vam} and references therein).

Here we will show that this Rainbow (energy-dependent) metric is not invariant under a deformed boost.  It should be noticed, in fact, that this class of metrics does not automatically leads to a flat invariant (under a ten-generator deformed Poincar\'{e} group) limit for the line element $ds^2$. Thus, Rainbow metric phenomenology may seem more suited to formalize LIV scenarios than (deformed) symmetric ones. However we will show how such energy-dependent metrics play an important role at the kinematical level in MDR-inspired Finsler geometries \cite{Girelli:2006fw,Amelino-Camelia:2014rga,VacaruPheno1,VacaruPheno2,VacaruPheno3}, and also in a maximally symmetric scenario. 

%Finsler geometry is basically Riemannian geometry without the quadratic restriction,

Finsler geometry is analogous to Riemannian geometry. However, a typical difference is that in Finsler geometry objects are defined on the tangent bundle while in Riemann geometry they live on $M$. Another important difference is that in Riemannian geometry there is a unique connection compatible with the metric as opposite to the Finsler case where one can have different possibilities \cite{Minguzzi:2014fxa}. This formalism has been proven to be very relyable in providing a powerful tool to investigate on non-standard particle physics and models of quantum-gravity on anisotropic spacetimes, see for instance \cite{VacaruQG1,VacaruQG2}. Finsler geometry makes possible to formalize a generalization of the relativistic Lagrangian formalism in the description of the kinematics of a single particle on curved momentum and spacetime geometries, with four-velocity-dependent metrics (an exploration on the Hamiltonian approach to such a framework can be found in \cite{Christian1,Christian2}). Another generalization of relativistic theories from a Hamiltonian approach can be formalized within the so called Relative Locality framework \cite{bob,principle,kbob,lateshift,SpecRelLoc}, in which the Hamiltonian is identified as an invariant element in a curved momentum-space. In this framework, the fundamental metric is the momentum-space one $\zeta^{\alpha\beta}(p)$. These approaches describing $\ell$-deformed theories do not contradict each other, but the two metrics play different roles in describing the kinematics of particles subject to MDR: the Finsler metric enters in the description of the Lagrangian formalism, and the momentum-space metric in the Hamiltonian one. Since these structures (from Rainbow, Finsler and momentum-space approaches) are symmetric, bilinear and non-degenerate maps, and are sufficient to find world-lines and dispersion relations (however using different methods), they can be properly defined as metrics. %In this paper we will describe in detail the features and complications related to the metric formalism of Planck-scale deformed theories, with particular attention to the massless case. In fact, the Finsler formalism introduced in reference \cite{Girelli:2006fw} presents a discontinuity in the limit $m\rightarrow 0$. 

For definitiveness, in this paper we will work with a MDR described by the generic Hamiltonian widely studied in the literature on a 1+1 dimensional expanding universe (see for instance \cite{Rosati:2015pga} and references therein): 
\begin{equation}
{\cal H}=a^{-2}(\eta)(\Omega^2-\Pi^2)+\ell a^{-3}(\eta)(\gamma \Omega^3+\beta \Omega \Pi^2)\,,\label{Hamiltonian}
\end{equation}
where $(\eta,x)$ are the so called conformal time coordinates, $(\Omega,\Pi)$ are their conjugate momenta, $a(\eta)$ is the scale factor of the universe, $\beta$ and $\gamma$ are two numerical parameters of order 1 and $\ell\sim 1/M_P$ is the deformation parameter, where $M_P\sim 1.2\times 10^{28}\, \mathrm{eV}$ is the Planck mass in units where $c=\hslash=1$. The MDR is recovered imposing the on-shell relation $\mathcal{H}=m^2$.

The paper is organized as follows: in Section \ref{sec:2} we mention some issues in the Rainbow metric approach which are relevant to the arguments presented in this paper. In section \ref{sec:3} we review the Lagrangian formalism and the role of momentum-space metrics; we will also give a glimpse on the symmetries for both scenarios, but a detailed study will be presented in \cite{Loboetalinpreparation2016}. Section \ref{sec:4} is devoted to a review of the MDR-inspired Finsler geometries introduced in \cite{Girelli:2006fw,Amelino-Camelia:2014rga}. Next, in section \ref{sec:5}, we discuss the problems found in the MDR-Finsler approach regarding the massless limit and propose an alternative way to handle it by re-writing the action for the particle as a Polyakov-like action. This approach allows us to obtain a metric that describes the same MDR kinematics and whose massless case is well defined as the continuous limit from the massive one.  A detailed description of these metrics can be found in \cite{Loboetalinpreparation2016}. In section \ref{sec:6} we discuss about particles' dynamics, geodesic equations and worldlines in the different formalisms described in the previous sections. It is important to annotate that the work presented here is only related to the kinematics of particles subject to MDR in a relativistic description using deformed symmetries. At this stage we do not pursue a fundamental theory, instead we aim for an effective description which eventually will allows us to make contact with a quantum gravity phenomenology of spacetime. There are different perspectives where a fundamental description of quantum gravity is proposed by using Finsler geometry and where the study of N-connections is fully justified, e.g. MDR in a LIV framework as the Horava-Lifshitz theory, see for example \cite{VacaruPheno3}. Therefore, our discussion on connections in Finsler geometry will be limited to the minimum required. It is worth to mention that all the results are valid up to first order in the deformation parameter $\ell$, but the technique may be straightforwardly applied to higher order perturbations, with the only requirement of having a well-defined Legendre transformation relating the Hamiltonian with the Lagrangian. Finally, in section \ref{sec:7} we give some closing remarks about the results here presented.

%%%%%%%%%%%%%%%%%%%%%%%%%%%%%%%%%%%%%%%%%%%%%%%%%%%%%%%%%%%%%%%%%%%%%%%%%%%%%%%%%%%%%%%%%%%%%%%%%%%%%%%%%%

\section{Deformed symmetries and Rainbow metrics}
\label{sec:2}

The purpose of this section is to clarify some aspects about the symmetries within the Rainbow metric approach. Consider the case $a(\eta)=1$ of the Rainbow line-element \cite{Rainbow} related to the Hamiltonian (\ref{Hamiltonian})
\begin{equation} \label{RainLineEl}
ds^2=(1-\ell \gamma p_0)(\d x^0)^2-(1+\ell \beta p_0)(\d x^1)^2\, .
\end{equation}
The Hamiltonian (\ref{Hamiltonian}) is invariant (in the flat spacetime limit) under a set of $\ell$-deformed Lorentz transformations. A deformed boost generated by
\begin{equation}  \label{BoostGen}
{\cal N}=x^0 p_1(1-\ell\gamma p_0)+x^1\left(p_0+(\beta+\frac{\gamma}{2})\ell p_0^2+\frac{\ell}{2}\beta p_1^2\right),
\end{equation}
has a finite action on an observable $A$ which can be expressed in terms of the Poisson brackets as
\begin{equation}
A'={\cal B}\rhd A \equiv A + \xi\{{\cal N},A\}+\frac{\xi}{2!}\{{\cal N},\{{\cal N},A\}\}+\ldots\, ,\label{BoostTransf}
\end{equation}
where $\xi$ is the rapidity parameter.\footnote{See \cite{Loboetalinpreparation2016} for more details on the representation of this boost.}
It can be shown that, despite having $\{{\cal N},{\cal H}\}=0$, the line-element (\ref{RainLineEl}) is not invariant, but at first order in the rapidity parameter $\xi$, transforms to
\begin{equation}
(ds^2)'=ds^2-\ell\xi(\beta p_1 (\d x^1)^2+\gamma p_1 (\d x^0)^2)\,.
\end{equation}
This non invariance poses a problem from a relativistic point of view, since the norm of vectors would not be invariant under a deformed transformation. Moreover, this changes the perspective of this working framework; since we cannot identify local invariant observers under deformed Poincar\'e transformations, it is necessary to break Lorentz invariance. This property has important consequences for the definition of photon's trajectories in Rainbow Gravity, since $ds^2=0$ does not define locally-invariant worldlines.

Therefore, Rainbow metrics seem to suit better a LIV-like phenomenology than a deformed-relativistic one. This, in turn, is related to the non-invariance under a boost (\ref{BoostGen}) of the $\ell$-deformed Lagrangian (see reference \cite{Amelino-Camelia:2014rga}). So, as long as a breaking of the Lorentz invariance is not ruled out, Rainbow metrics could be a useful approach to cosmological LIV-phenomenology.

%%%%%%%%%%%%%%%%%%%%%%%%%%%%%%%%%%%%%%%%%%%%%%%%%%%%%%%%%%%%%%%%%%%%%%%%%%%%%%%%%%%%%%%%%%%%%%%%%%%%%%%%%%

\section{Lagrangian formalism and momentum-space metrics}
\label{sec:3}
%In order to describe the Lagrangian formalism and analyze its symmetries we set the scale factor to $a(\eta) = (1 - H \eta)^{-1}$, which describes a constant rate expanding (de Sitter) universe. 
From the Hamiltonian (\ref{Hamiltonian}) it is possible to write the action
	\begin{equation}  \label{h-action}
	{\cal S}[q,p,\lambda] = \int d\tau \big[\dot{\eta}\, \Omega + \dot{x}\, \Pi-\lambda ({\cal H} - m^2)\big],
	\end{equation}
where $\lambda$ is introduced as a Lagrange multiplier to enforce the mass-shell condition; $q^\mu = (\eta, x)$, $p_\mu = (\Omega, \Pi)$ are the spacetime and momentum space coordinates and $\dot{q}\equiv dq/d\tau$.
Using Hamilton equations
\begin{equation}
\dot{q}^\alpha=\lambda\{\mathcal{H},q^\alpha\}\,,\label{HamiltEq}
\end{equation}
we can express the action (\ref{h-action}) in terms of the four-velocities $\dot{q}^\alpha$ 
\begin{eqnarray}\label{p-action}
{\cal S}[q,\lambda] &=& \int d\tau {\cal L}(q,\dot{q},\lambda) =
 \int d\tau\Big(\frac{a^2(\eta)}{4\lambda}(\dot{\eta}^2-\dot{x}^2)+\nonumber\\
 &-& \frac{\ell a^3(\eta)}{8\lambda^2}(\gamma\dot{\eta}^3+\beta\dot{\eta}\dot{x}^2) +\lambda m^2\Big).
\end{eqnarray}

If $m \neq 0$, it is possible to solve for the Lagrange multiplier $\lambda = \lambda(q, \dot{q})$ from the extremization of the action, $\delta S/\delta \lambda = 0$ 
	\begin{equation}\label{lambda}
	\lambda=\frac{1}{2m}a(\eta)\sqrt{\dot{\eta}^2-\dot{x}^2}-\frac{\ell}{2}\frac{a(\eta)(\gamma\dot{\eta}^3+\beta\dot{\eta}\dot{x}^2)}{\dot{\eta}^2-\dot{x}^2}.
	\end{equation}
Substituting $\lambda$ into (\ref{p-action}) give us the Lagrangian depending on coordinates and velocities ${\cal L}(q,\dot{q})$:
	\begin{equation}  \label{lagmqdotq}
	\mathcal{L}(q,\dot{q}) = m a(\eta) \sqrt{\dot{\eta}^2 - \dot{x}^2} - \frac{\ell}{2}\, m^2 a(\eta)\, \frac{\beta \dot{\eta} \dot{x}^2 + \gamma \dot{\eta}^3}{\dot{\eta}^2 - \dot{x}^2}.
	\end{equation}
At first order in the deformation parameter $\ell$ it is possible to express the Lagrangian (\ref{lagmqdotq}) as
\begin{equation}
\mathcal{L}(q,\dot{q}) = m\sqrt{g_{\mu\nu}(q,\dot{q})\dot{q}^\mu\dot{q}^\nu}\,,\label{LagMetr}
\end{equation}
where $g_{\mu\nu}(q, \dot{q})$ can be identified, as we will see in the next section, as a four-velocity-dependent spacetime metric within the Finsler formalism. Inverting the relation between four-velocities and four-momenta, it is possible to think of $g_{\mu\nu}$ as momentum dependent. This metric is, in general, not invariant under the deformed set of symmetries of the MDR defined by $\mathcal{H} = m^2$ \cite{Amelino-Camelia:2014rga,Mignemi}, and in this sense we can regard it as some sort of Rainbow metric.

\subsection{Momentum-space metric}

As we mentioned before, in the context of Relative Locality, momentum space is curved and the metric for this space allows to interpret the Planck-scale DSR as a spacetime manifestation of momentum-space curvature. In this framework, unusual features like energy-dependent time delays and deformed composition laws can be interpreted as dual redshift effects and composition laws in a curved manifold \cite{lateshift}. 
In the Relative Locality framework, the mass-shell relation is obtained as the geodesic distance (from the momentum space origin to the particle's momentum) of the momentum-space metric $\zeta$ as
\begin{equation}  \label{Ham_momentum_metric}
{\cal H}= m^2 =\int_0^1\zeta^{\mu\nu}(p)\dot{p}_\mu \dot{p}_\nu\,d\sigma\,,
\end{equation}
where $\dot{p}_\mu(\sigma)$ is the tangent vector to the momentum-space geodesics parametrized by $\sigma$.\footnote{A systematic discussion on this topic can be found in \cite{GiuPalmisano}.} In our case the momentum space metric (in the Minkowskian limit in 1+1dimensions) can be represented by the diagonal matrix
\begin{equation}
\zeta^{\mu\nu}=\left(\begin{array}{cc}
1+2\gamma\ell p_0 & 0\\
0 & -(1-2\beta\ell p_0)
\end{array}\right)\,.\label{mommetrexpl}
\end{equation}
 Interestingly, the Hamiltonian $\mathcal{H}$ and the Lagrange multiplier $\lambda$ (\ref{lambda}), can be expressed as the algebraic relations
\begin{equation}
{\cal H}=\frac{\zeta_{\mu\nu}(\dot{q})\dot{q}^\mu\dot{q}^\nu}{4\lambda^2}\;,\;\;\;\lambda=\frac{\sqrt{\zeta_{\mu\nu}(\dot{q})\dot{q}^\mu\dot{q}^\nu}}{2m}\,, \label{HameLamb}
\end{equation} 
where $\zeta_{\mu\nu}$ are the components of the momentum-space metric in terms of the four-velocity. These algebraic relations may lack of a geometrical meaning, nevertheless suggest the relevance of the momentum-space metric in the relativistic description of the DSR kinematics.

Observing equation (\ref{mommetrexpl}) and using (\ref{HameLamb}) we notice that from the case $a(\eta)\equiv1$ we can recover a simple expression for the (deformed) special relativity spacetime norm, defining spacetime line-element as
\begin{equation}
ds^2\hspace{-0.1cm}=\zeta_{\mu\nu}\d x^\mu \d x^\nu\hspace{-0.1cm}=\hspace{-0.1cm}(1-2\ell \gamma p_0)(\d x^0)^2-(1+2\ell \beta p_0)(\d x^1)^2.\label{RelLocLineEl}
\end{equation}
We can repeat what we did with (\ref{RainLineEl}) to show that however in this case this line-element is indeed invariant:
\begin{equation} \label{lineElemInv}
(ds^2)'=ds^2\,.
\end{equation}
Therefore, in 3+1D this formalism allows us to define a class of locally flat observers immersed in a 10 $\ell$-deformed generators symmetric spacetime, formalized within the coherent framework of special Relative Locality \cite{SpecRelLoc}. This metric is however not sufficient to express all the general relativistic features we need to describe particles motion in Planck-scale curved spacetime, like for instance connections and Killing vectors. In order to add those further elements to our picture we need in fact to deepen more in MDR realization in Finsler geometry. 

%%%%%%%%%%%%%%%%%%%%%%%%%%%%%%%%%%%%%%%%%%%%%%%%%%%%%%%%%%%%%%%%%%%%%%%%%%%%%%%%%%%%%%%%%%%%%%%%%%%%%%%%%%

\section{MDR-inspired Finsler geometries}
\label{sec:4}

In \cite{Girelli:2006fw} it was pointed out that the Lagrangian (\ref{LagMetr}) can be identified with a MDR-related Finsler norm $F(\dot{q})$, that is, the Lagrangian can be expressed as 
\begin{equation}
\mathcal{L}(q,\dot{q}) = m F(\dot{q})\, .\label{Fins1}
\end{equation}
It can be straightforwardly verified that the $F(\dot{q})$ related to (\ref{Hamiltonian}) satisfies the conditions of positivity and homogeneity:
\begin{equation}
\left\{\begin{array}{l}
   F(\dot{q})\neq 0\;\;\; {if}\;\;\; \dot{q}\neq 0\\
   F(\epsilon\dot{q})=|\epsilon|F(\dot{q}) \label{FinslerConditions}
\end{array}\right. \,.
\end{equation}
Therefore, the Finsler metric $g^{{\rm F}}(q,\dot{q})$ can be defined, accordingly to the metric in (\ref{LagMetr}), just imposing its components to be homogeneous functions of degree zero, resulting in metric components which are proportional to the Hessian of the squared Finsler norm
\begin{equation}\label{finslermetric}
g_{\mu\nu}^{{\rm F}}=\frac{1}{2}\frac{\partial^2 F^2}{\partial\dot{q}^{\mu}\partial\dot{q}^{\nu}}\,.
\end{equation} 
This metric satisfies the relations
\begin{equation}\label{partial_hom}
\dot{q}^{\alpha}\frac{\partial g_{\mu\nu}^{{\rm F}}}{\partial \dot{q}^{\alpha}}=\dot{q}^{\mu}\frac{\partial g_{\mu\nu}^{{\rm F}}}{\partial \dot{q}^{\alpha}}=\dot{q}^{\nu}\frac{\partial g_{\mu\nu}^{{\rm F}}}{\partial \dot{q}^{\alpha}}=0,
\end{equation}
which allow to write the equations of motion from the extremization of the arc-length (geodesic equations) as \footnote{Property (\ref{partial_hom}) results in the cancellation of a considerable amount of terms that appear in the geodesic equations simplifying a significantly their appearance.}
\begin{equation}
\ddot{q}^{\alpha}+\Gamma^{\alpha}_{\mu\nu}(q,\dot{q})\dot{q}^{\mu}\dot{q}^{\nu}=0\,,\label{Geodesic0}
\end{equation} 
where the coefficients $\Gamma^\alpha_{\mu\nu}$ have the usual form of the Christoffel symbols in terms of the derivatives of the metric with respect to the coordinates, but keeping an explicit dependence on the four-velocity. These equations \eqref{Geodesic} describe the worldlines of massive particles subject to a MDR and coincide with the those obtained solving the Hamilton equations subject to the mass-shell condition.

An important contribution of the Finsler approach to this framework is undoubtedly the possibility to define a deformed Killing equation. In fact, assuming a metric to be $\dot q$-dependent, in the flat space-time case, one easily obtains
\begin{equation}
g_{\alpha\nu}\partial_\mu \xi^\alpha+g_{\mu\alpha}\partial_\nu \xi^\alpha+\frac{\partial g_{\mu\nu}}{\partial \dot{q}^\beta}\partial_\alpha\xi^\beta\dot{q}^\alpha=0\, ,
\end{equation}
from which it is possible to obtain the boost generator \eqref{BoostGen}. The same equation holds both for the Finsler and the Rainbow approach.\\
So far we have observed that, in the geometric formalization of Planck-scale MDRs, at least two metrics come into play: a momentum-space metric $\zeta$ which allows an invariant description of the physics of locally-flat observers and a spacetime (Finsler) metric $g$ whose geodesics are the worldlines of the particles.
Here again, we find intriguing relations between these metrics 
\begin{eqnarray}
\frac{\zeta_{\mu\nu}(\dot{q})\dot{q}^\mu\dot{q}^\nu}{4\lambda^2} &=& g^{\alpha\beta}(p)p_\alpha p_\beta\,,\label{dual1}\\
\frac{g_{\mu\nu}(\dot{q})\dot{q}^\mu\dot{x}^\nu}{{4\lambda^2}} &=& \zeta^{\alpha\beta}(p)p_\alpha p_\beta  \label{dual2}\,.
\end{eqnarray} 
It is important to notice that on the left hand side of equations \eqref{dual1} and \eqref{dual2} we have tensorial objects, on the other hand, on the right side we have algebraic relations involving the components of the metrics. It would be interesting to find a unified framework where these two metrics are a manifestation of a single geometrical object.

\subsection{Aside comment on connections in Finsler geometry}

In the flat spacetime limit \cite{Amelino-Camelia:2014rga} a $\kappa$-Poincar\'{e}-inspired MDR model can be described as a Berwald space in which the connection does not depend on spacetime coordinates.
\par

Equations \eqref{Geodesic0} can be written as
\begin{equation}
g_{\gamma\sigma}^F(q,\dot{q})\ddot{q}^{\sigma}+\left(\frac{\partial g_{\alpha\gamma}^F(q,\dot{q})}{\partial q^\sigma}-\frac{1}{2}\frac{\partial g_{\alpha\sigma}^F(q,\dot{q})}{\partial q^\gamma}\right)\dot{q}^{\alpha}\dot{q}^{\sigma}=0\,,\label{Geodesic}
\end{equation}  
where we can identify the spray coefficients
\begin{equation}
G^\alpha(q,\dot{q})\hspace{-0.1cm}=\hspace{-0.1cm}\frac{g^{\alpha\beta}_F(q,\dot{q})}{2}\left(\frac{\partial g_{\beta\gamma}^F(q,\dot{q})}{\partial q^\mu}-\frac{1}{2}\frac{\partial g_{\gamma\mu}^F(q,\dot{q})}{\partial q^\beta}\right)\dot{q}^\gamma \dot{q}^\mu\,.
\end{equation}
\par
Equations \eqref{Geodesic} describe the worldlines of massive particles subject to a MDR and coincide with those obtained solving the Hamilton equations subject to the mass-shell condition.
In \cite{LetiziaLiberatiinprep} the curved spacetime case have been studied, finding that only in a very special limit, $\ell\neq 0\, ,H\neq 0\, , \ell H\rightarrow 0$ (where H is the Hubble constant), such a model can still be considered a Berwald space.
\par
In Finsler geometry one deals with tensors on the tangent bundle (sometimes called d-tensors), so it is useful to introduce a non-linear connection $N$ to split the tangent space to the tangent bundle in horizontal and vertical spaces, which in turn allows to define a covariant derivative and the notion of parallel transport. This splitting is characterized by the coefficients $N^\alpha_\beta$ which allow to define a frame field for the tangent spaces to the tangent bundle as
\begin{equation}
\delta_{\alpha} \doteq \frac{\partial}{\partial q^\alpha}-N^\beta_\alpha\frac{\partial}{\partial y^{\beta}} \qquad e_\gamma \doteq \frac{\partial}{\partial y^{\gamma}}\,,
\end{equation}
where $(q^{\alpha},y^{\gamma} = \dot{q}^{\gamma})$ are local coordinates on the tangent bundle.
\par
Given an spray $G$, there is a connection $N$ whose spray is $G$, defined by
\begin{equation}
N^\alpha_\beta = \frac{\partial G^\alpha}{\partial y^\beta}\, ,
\end{equation}
for which the paths of the spray coincide with the geodesics for the connection.

A Finsler connection is a pair $(N,\nabla)$ where $N$ is a nonlinear connection on the tangent bundle and $\nabla$ a linear connection on the vertical space. Then a Finsler connection is determined locally by the coefficients \\$(N^\alpha_\beta, \textrm{G}^{\alpha}_{\mu\nu}, C^{\alpha}_{\mu\nu})$ where $\textrm{G}^{\alpha}_{\mu\nu}$ and $C^{\alpha}_{\mu\nu}$ are collections of locally defined homogeneous functions of degree 0 with appropriate transformation rules \cite{dahl}. In this case, $\textrm{G}^{\alpha}_{\mu\nu}$ and $C^{\alpha}_{\mu\nu}$ are the coefficients of the linear connection for derivatives in the direction of the basis vectors of the horizontal and vertical spaces respectively, $(\delta_\alpha, e_{\gamma})$. Let 
\begin{eqnarray}
%{\cal G}^{\alpha}_{\beta}&\doteq&\partial G^{\alpha}(q,y))/\partial y^{\beta},\\
\textrm{G}^{\alpha}_{\mu\nu}&\doteq&\partial^2 G^{\alpha}(q,y)/\partial y^{\mu}\partial y^{\nu},\\
\Gamma^{\alpha}_{\mu\nu}&\doteq&\frac{1}{2}g_F^{\alpha\beta}\left(\delta_{\mu}g^F_{\nu\beta}+\delta_{\nu}g^F_{\mu\beta}-\delta_{\beta}g^F_{\mu\nu}\right),\\
C^{\alpha}_{\mu\nu}&\doteq&\frac{1}{2}g_F^{\alpha\beta}\frac{\partial g^F_{\beta\mu}}{\partial y^\nu}\,.
\end{eqnarray}
Some notable Finsler connections are \cite{Minguzzi:2014fxa}
\begin{eqnarray}
&\textrm{Berwald:}& \ \ (N^{\alpha}_{\beta},\textrm{G}^{\alpha}_{\mu\nu},0),\label{finslerconnection1}\\
&\textrm{Cartan:}& \ \ (N^{\alpha}_{\beta},\Gamma^{\alpha}_{\mu\nu},C^{\alpha}_{\mu\nu}),\label{finslerconnection2}\\
&\textrm{Chern-Rund:}& \ \ (N^{\alpha}_{\beta},\Gamma^{\alpha}_{\mu\nu},0),\label{finslerconnection3}\\
&\textrm{Hashiguchi:}& \ \ (N^{\alpha}_{\beta},\textrm{G}^{\alpha}_{\mu\nu},C^{\alpha}_{\mu\nu}).\label{finslerconnection4}
\end{eqnarray}
\par
One can verify that due to the validity of (\ref{partial_hom}) (which is a direct consequence of the definition of the metric as the Hessian of a 2-homogeneous function (\ref{finslermetric})), then the autoparallel curves defined from the above Finsler connections coincide with the extremizing geodesics (\ref{Geodesic}). Therefore, a pure kinematical analysis of the geodesics would not permit to distinguish between these proposals.
\par 
We should however anticipate that in the case here under scrutiny (a photon with deformed Hamiltonian (\ref{Hamiltonian}) propagating in an expanding spacetime) the Finsler formalism cannot be completely applied, and we will need to consider a generalized case and a slightly different approach will be adopted in order to study the Euler-Lagrange equations and derive the particles worldlines.

In the following section we will see that for this MDR-Finsler metric the massless limit is not well defined and that, renouncing to the properties formalized in \eqref{partial_hom}, a spacetime metric can be defined for which the limit $m \to 0$ presents no complications and properly describes the particle's dynamics.

%%%%%%%%%%%%%%%%%%%%%%%%%%%%%%%%%%%%%%%%%%%%%%%%%%%%%%%%%%%%%%%%%%%%%%%%%%%%%%%%%%%%%%%%%%%%%%%%%%%%%%%%%%

\section{The massless limit and the spacetime metric from a Polyakov-like action}
\label{sec:5}

From the expression \eqref{LagMetr} and the Finsler metric in \cite{Amelino-Camelia:2014rga}, it seems that the massless case cannot be handled within the MDR-Finsler approach not even in the $a(\eta) = 1$ case, even though the description from the action \eqref{p-action} does not present inconsistencies when $m=0$ and using Hamiltonian dynamics the massless case can be completely solved.

In the massive case the Lagrange multiplier $\lambda$ was determined by the extremization of the action $\delta S/\delta \lambda = 0$ yielding to \eqref{lambda}, however in the massless case the on-shell relation written in terms of the four-velocities
	\begin{equation}
\frac{a^2(\eta)}{4\lambda^2}(\dot{\eta}^2-\dot{x}^2)-2\,  \frac{\ell a^3(\eta)}{8\lambda^3}(\gamma\dot{\eta}^3+\beta\dot{\eta}\dot{x}^2)=0\,,\label{lambnonmass}
	\end{equation}
does not provide any information about $\lambda$. Notice that Eq. (\ref{lambnonmass}) presents a factor of $2$ on its second term, which is different from the Lagrangian of (\ref{p-action}). %\footnote{At first glimpse the solution that one obtains from \eqref{lambnonmass} seems to suggest $\lambda\sim f(\dot{q})/\ell$, which would be incompatible with our formalism at first order in $\ell$. Anyways if we had $f(\dot{q})\sim \ell$, then it would be possible to obtain reasonable physical solutions, preserving the Finsler formalization of the theory (more details on this can be found in Ref.\cite{LetiziaLiberatiinprep}).}

%One could think of a workaround by fixing the Lagrange multiplier to $\lambda=1/2$, but that would mean to loose reparametrization invariance, therefore we will not pursue this option in this work (which will be explored in detail in \cite{Loboetalinpreparation2016}). 

In the undeformed case ($\ell=0$) finding $\lambda$ from the on-shell relation and \eqref{h-action}, guarantees that the action
	\begin{equation} \label{unNG}
	{\cal S}[q]=m\int d\tau\sqrt{g_{\mu\nu}(q)\, \dot{q}^{\mu}\dot{q}^{\nu}}\, ,
	\end{equation}
is invariant under reparametrizations. Nevertheless, it is also possible to write the action in a classically equivalent way leaving the extra degree of freedom $\lambda$ unspecified
	\begin{eqnarray}  \label{unPol}
	{\cal S}[q,\lambda] =  \int d\tau \bigg(\frac{1}{4\lambda}g_{\mu\nu}(q)\, \dot{q}^\mu \dot{q}^\nu +\lambda m^2 \bigg).
	\end{eqnarray}
These two actions are equivalent since they give rise to the same equations of motion with the bonus that \eqref{unPol} is invariant under reparametrizations. %even in the massless case $m=0$
The latter action can be identified as a Polyakov-like version of the former which in turn, can be thought as a Nambu-Goto-like action.\footnote{Interestingly, Refs.\cite{VacaruNambu1,VacaruNambu2} develop a (super) string approach to gravity in the Finsler scenario.}

For the deformed case we can find a Polyakov-like expression from which we can obtain in a systematic way a four-velocity-dependent spacetime metric.

Since the Finsler metric derived from the Nambu-Goto-like action, which can be identified with the arc-length function, is not well-defined in the massless limit, we will find convenient to derive a metric from the Polyakov-like version of the single particle action. 

The solution to this discontinuity problem between massive and massless particles could be of highest relevance since particles with very small, but finite masses (e.g. neutrinos), may be described as being massive or massless depending on the role that their masses play into the phenomenological model. It is useful then to be able to rely on a single comprehensive formalism.

The derivation of a Finsler metric from the arc-length functional is a long time studied formalism, for which exist an extensive literature (see \cite{HRund} and references therein). This procedure was also used in Refs. \cite{Girelli:2006fw,Amelino-Camelia:2014rga} to derive the metric probed by massive particles, given that the arc-length is the action for these kind of particles. 

In this paper, the objective of our Polyakov-like approach is to propose an alternative to Girelli-Liberati-Sindoni's \cite{Girelli:2006fw} Finsler metric coming from a dispersion relation. In their paper, the Finsler approach serves to provide a rigorous realization of Rainbow metrics, since, in their words {\it ``it involves a metric defined in the tangent bundle, while depending on a quantity associated to the cotangent bundle (i.e. the energy).''}

Although solving that issue, their proposed metric is not well-behaved in the massless limit. Henceforth, we propose an alternative way of defining an object that fulfills the definition of a metric tensor and can be defined from an action functional that is well-defined for both massive and massless cases which, in fact, unifies them. Despite escaping the standard Finsler geometry approach, our metric still presents some properties of the previous case, like a parametrization-invariant and four-velocity-dependent metric, besides reproducing the dispersion relation from a norm.
%we propose alternately, a physical way of defining a unique four-velocity-dependent and parametrization invariant spacetime metric that would be probed by any particle, massless or massive, from the Polyakov-like action.  

The simplest approach in the search for uniqueness is to realize that the integrand of \eqref{p-action} is an analytic function and can be expressed as Taylor expansion in the velocities 
\begin{eqnarray}
{\cal S}[q,\lambda] &=& \int d\tau\bigg[{\cal L}\big|_{\dot{q} = 0} + \frac{\partial {\cal L}}{\partial 	\dot{q}^{\mu}}\bigg|_{\dot{q} = 0}\dot{q}^{\mu} +\frac{1}{2!}\frac{\partial^2 {\cal L}}{\partial \dot{q}^{\mu}\partial \dot{q}^{\nu}}\bigg|_{\dot{q} = 0}\dot{q}^{\mu}\dot{q}^{\nu}\hspace{-0.1cm}+\nonumber\\
	&+&\frac{1}{3!}\frac{\partial ^3 {\cal L}}{\partial 	\dot{q}^{\mu}\partial\dot{q}^{\nu}\partial\dot{q}^{\gamma}}\bigg|_{\dot{q} = 0}\dot{q}^{\mu}\dot{q}^{\nu}\dot{q}^{\gamma} +...+\lambda m^2 \bigg],
	\end{eqnarray}
where the zeroth and first order terms vanish as well as those of higher than the third order. The action can be then expressed as 
	\begin{equation}
	{\cal S}[q,\lambda]=\int d\tau \left[\frac{1}{4\lambda}\, \tilde{g}_{\mu\nu}(q,\dot{q},\lambda)\, \dot{q}^{\mu} \dot{q}^{\nu} + \lambda 	m^2\right],
	\end{equation}
for which
	\begin{equation} \label{metricg}
	\tilde{g}_{\mu\nu}(q,\dot{q}, \lambda) = g^0_{\mu\nu}(q) + g^1_{\mu\nu}(q, \dot{q}, \lambda)\,,
	\end{equation}
and where we have identified
	\begin{equation}
	\frac{1}{4\lambda} g^0_{\mu\nu}=\frac{1}{2!}\frac{\partial^2 {\cal L}}{\partial \dot{q}^{\mu}\partial \dot{q}^{\nu}}\bigg|_{\dot{q} = 0}\;, \; \frac{1}{4\lambda} g^1_{\mu\nu} = \frac{1}{3!}\frac{\partial ^3 {\cal L}}{\partial \dot{q}^{\mu} \partial \dot{q}^{\nu} \partial\dot{q}^{\gamma}}\bigg|_{\dot{q} = 0}\dot{q}^{\gamma}.
	\end{equation}
This four-velocity-dependent metric $g_{\mu\nu}(q,\dot{q},\lambda)$ encloses the massive and massless particle cases through its dependence on $\lambda$. Notice that this is not a standard Finsler metric, not even in the $m \neq 0$ case as it is not defined as the Hessian of the squared Finsler function. In the following subsections we present the massive and massless particle cases, for which the limit $m \to 0$ can be consistently taken.

%%%%%%%%%%%%%%%%%%%%%%%%%%%%%%%%%%%%%%%%%%%%%%%%%%%%%%%%%%%%%%%%%%%%%%%%%%%%%%%%%%%%%%%%%%%%%%%%%%%%%%%%%%

\subsection{Massive and massless particles}

For the massive particle case we can define the four-velocity-dependent metric as in \eqref{metricg} and using the expression for $\lambda$  given in \eqref{lambda} we can write the action as follows
	\begin{equation}  \label{actionmassive}
	{\cal S}[q]= m \int d\tau \sqrt{\tilde{g}_{\mu\nu}\dot{q}^{\mu}\dot{q}^{\nu}},
	\end{equation}
where
\begin{equation} \label{massiveg}
	\tilde{g}_{\mu\nu} =a^2(\eta)
	\left(\begin{array}{cc}
	1 - \ell\,  \gamma \frac{m\dot{\eta}}{\sqrt{\dot{\eta}^2-\dot{x}^2}} & -\frac{1}{3}\ell \beta\frac{m\dot{x}}{\sqrt{\dot{\eta}^2-\dot{x}^2}}\\
	-\frac{1}{3}\ell \beta\frac{m\dot{x}}{\sqrt{\dot{\eta}^2-\dot{x}^2}} &  -1 - \frac{1}{3}\ell\, \beta\frac{m\dot{\eta}}{\sqrt{\dot{\eta}^2-\dot{x}^2}} 
	\end{array}\right)\,.
\end{equation}

The extremization of this action furnishes the equations of motion of massive particles.

\par
In the case of a massless particle it is not possible to find a definite solution for $\lambda$ as it is done in the $m \neq 0$ case. This can be seen from the equations of motion, however, the solution $x(\eta)$ for the particle's world-line is obtained independently from $\lambda$. Therefore, we can absorb $\lambda$ in the re-parametrization $s(\tau)$ such that $2\lambda d\tau=ds$ and the action takes the standard form
	\begin{equation} \label{action-s}
	{\cal S}[q]=\frac{1}{2}\int ds\, \tilde{g}_{\mu\nu}(q,q')q'^{\mu}q'^{\nu},
	\end{equation}
where $q'\equiv dq/ds$. The extremization of this action \mbox{$\delta {\cal S} = 0$} along with the on-shell condition ${\cal H}=0$ written in terms of the velocities $q'$ furnishes the equations of motion whose solutions describe the trajectory for the massless particle.

%%%%%%%%%%%%%%%%%%%%%%%%%%%%%%%%%%%%%%%%%%%%%%%%%%%%%%%%%%%%%%%%%%%%%%%%%%%%%%%%%%%%%%%%%%%%%%%%%%%%%%%%%%

\subsection{Energy-momentum-dependent metric}

From the Lagrangian in \eqref{h-action}, the conjugate momenta to $q$ are $\partial {\cal L}/\partial \dot{q}$, that is
	\begin{eqnarray} 
	\Omega& = &a^2(\eta)\frac{\dot{\eta}}{2\lambda} - \frac{\ell a^3(\eta)}{8\lambda^2}(3\gamma\dot{\eta}^2\,+\,\beta\dot{x}^2) \,,\label{conjmom1}\\
	\Pi& = &-a^2(\eta)\frac{\dot{x}}{2\lambda}-\frac{\ell a^3(\eta)}{4\lambda^2}\beta\dot{\eta}\dot{x}.\label{conjmom2}
	\end{eqnarray}
This allows us to express the metric \eqref{metricg}, up to the first order in $\ell$, in terms of the energy and momentum of the particle as	
	\begin{equation}  \label{metricEM}
	\tilde{g}_{\mu\nu} = a^2(\eta)
	\left(\begin{array}{cc}
	1-\ell \gamma\Omega/a(\eta) & \frac{1}{3}\ell\, \beta\Pi/a(\eta)\\
\frac{1}{3}\ell \beta\Pi/a(\eta) & -1-\frac{1}{3}\ell\, \beta\Omega/a(\eta) 
	\end{array}\right)\,.
	\end{equation}
This metric describes both cases, that is, massive and massless particles and the limit $m \to 0$ from the massive case is well-defined through relations \eqref{conjmom1} and \eqref{conjmom2}. Its contravariant version when contracted with the conjugate momentum co-vector $P = p_\mu dq^\mu$ reproduces the Hamiltonian that describes the particle's dynamics
\begin{equation}  \label{metricHam}
\tilde{g}^{\mu\nu}p_{\mu}p_{\nu}=\frac{(\Omega^2-\Pi^2)}{a^2(\eta)}+\ell\frac{(\gamma \Omega^3+\beta \Omega \Pi^2)}{a^3(\eta)}={\cal H}\, ,
\end{equation}
fulfilling the Rainbow approach main assumption.

%%%%%%%%%%%%%%%%%%%%%%%%%%%%%%%%%%%%%%%%%%%%%%%%%%%%%%%%%%%%%%%%%%%%%%%%%%%%%%%%%%%%%%%%%%%%%%%%%%%%%%%%%%

\section{Worldlines and Geodesic equations}\label{sec:6}
In General Relativity the Levi-Civita parallel transport gives rise to free particles motion in the space manifold. If no force acts on the particle, so that it moves freely along a timelike path, we expect
the four-velocity to coincide at all times with itself. In other words we are asking the covariant derivative of $\dot{q}^\alpha$ to be zero:
\begin{equation}
\ddot{q}^\alpha+\Gamma^\alpha_{\beta\gamma}\dot{q}^\beta\dot{q}^\gamma=0\,.\label{geo0}
\end{equation}
%\begin{equation}
%\left\{\begin{array}{l}
%\ddot{\eta}=-\frac{\dot{a}}{a}(\dot{\eta}^2+\dot{x}^2)\,,\\
%\\
%\ddot{x}=-2\frac{\dot{a}}{a}\dot{\eta}\dot{x}\,.
%\end{array}\right.
%\end{equation}
In general solving this set of differential equations is rather difficult and, in order to obtain timelike geodesics it is often simplest to start from the spacetime metric, after dividing by $ds^{2}$ to obtain the form $g_{\mu\nu}\dot{q}^\mu\dot{q}^\nu=1$ or $g_{\mu\nu}\dot{q}^\mu\dot{q}^\nu=0$ in the massless case. This method has the advantage of bypassing a tedious calculation of Christoffel symbols.\\
This is true {\it a fortiori} in Rainbow Gravity where the Euler-Lagrange equation for a massless particle  with Lagrangian $\mathcal{L}=\frac{1}{2}g_{\mu\nu}(q,\dot{q})\dot{q}^\mu\dot{q}^\nu$,
\begin{equation}
\frac{d}{d\tau}\left(\frac{\partial {\cal L}}{\partial \dot{q}^\mu}\right)-\frac{\partial {\cal L}}{\partial q^\mu}=0\,,\label{EuLag}
\end{equation} 
defines a deformed version for the geodesic equation \eqref{geo0}. In fact since now the metric depends explicitely by $\dot{q}$ the \eqref{EuLag} becomes
\begin{equation}
\ddot{q}^\beta+\Gamma^\beta_{\rho\mu}\dot{q}^\rho\dot{q}^\mu +\Delta^\beta_{\rho\mu}\ddot{q}^\rho\dot{q}^\mu +E^\beta_{\rho\mu\nu}\dot{q}^\rho\dot{q}^\mu\dot{q}^\nu +Z^\beta_{\rho\mu\nu}\ddot{q}^\rho\dot{q}^\mu\dot{q}^\nu=0\,,\label{gen_geod1}
\end{equation}
in which
\begin{equation}
\left\{\begin{array}{l}
\Gamma^\beta_{\rho\mu}=\frac{1}{2}g^{\alpha\beta}(\partial_\rho g_{\mu\alpha}+\partial_\mu g_{\rho\alpha} -\partial_\alpha g_{\mu\rho})\\
\Delta^\beta_{\rho\mu}=g^{\alpha\beta}\left(\frac{\partial g_{\mu\alpha}}{\partial\dot{q}^\rho}+\frac{\partial g_{\rho\mu}}{\partial\dot{q}^\alpha}\right)\\
E^\beta_{\rho\mu\nu}=\frac{1}{2}g^{\alpha\beta}\frac{\partial^2 g_{\mu\nu}}{\partial q^\rho \partial \dot{q}^\alpha}\\
Z^\beta_{\rho\mu\nu}=\frac{1}{2}g^{\alpha\beta}\frac{\partial^2 g_{\mu\nu}}{\partial \dot{q}^\rho \partial \dot{q}^\alpha}
\end{array}\right.\,.
\end{equation}
Using Finsler formalism in this case does not simplify the solution of those differential equations, since in this formalism even if the geodesic equation in terms of the metric is classical, see Eq.\eqref{Geodesic}, the explicit equations, once that one writes down the metric in terms of its components, are exactly the same.\footnote{Smart solutions to find the worldlines expression in MDR-inspired Finsler formalism without solving the geodesic equations have however been found in \cite{Amelino-Camelia:2014rga,Loboetalinpreparation2016}.} In fact also equation \eqref{gen_geod1} can be re-expressed in the same form of the classical geodesic equation
\begin{equation}
\ddot{q}^{\rho}+\mathcal{G}^{\rho}_{\alpha\mu}(\dot{q})\dot{q}^{\alpha}\dot{q}^{\mu}=0\, ,
\end{equation}
in which ${\cal G}^{\rho}_{\alpha\mu}(\dot{q})\equiv {\cal Q}^{\rho}_{\beta}\Gamma^{\beta}_{\alpha\mu}+{\cal Q}^{\rho}_{\beta}E^{\beta}_{\alpha\mu\nu}\dot{q}^{\nu}$ and where, at first order in $\ell$, ${\cal Q}^{\rho}_{\gamma}\simeq\delta^{\rho}_{\gamma}-\Delta^{\rho}_{\gamma\mu}\dot{q}^{\mu}-Z^{\rho}_{\gamma\mu\nu}\dot{q}^{\mu}\dot{q}^{\nu}$. However rephrasing the differential equations in a different form does not decrase the complexity of their explicit expression.
\par
Interestingly, the metric defined in the previous section is a generalized Finsler metric \cite{krupkova}, for which it is also possible to identify a spray from\\ $G^{\rho}={\cal G}^{\rho}_{\alpha\mu}\dot{q}^{\alpha}\dot{q}^{\mu}$. The major difference with respect to the previous Finsler approach is the non-validity of identities (\ref{partial_hom}). In the generalized case we only have a $0$-homogeneous metric,\footnote{This class of geometries was studied in \cite{shimada} and references therein.}
\begin{equation}
\tilde{g}_{\mu\nu}(q,\epsilon\dot{q})=\tilde{g}_{\mu\nu}(q,\dot{q}) \, ; \, \epsilon>0.
\end{equation}
which simply implies that
\begin{equation}
\dot{q}^{\alpha}\frac{\partial \tilde{g}_{\mu\nu}}{\partial \dot{q}^{\alpha}}=0.
\end{equation}
\par
Therefore, for all of the above cited connections (\ref{finslerconnection1})-(\ref{finslerconnection4}) the autoparallel curves are not in general geodesics (that is only the case for the Berwald connection), as was the case of Finsler geometry. Then a more precise investigation is required to determine whether a Berwald connection may or may not be a compelling candidate in such generalized Finsler space.\footnote{The Berwald connection is defined from the spray coefficients of the geodesic equation, therefore the autoparallel curves are automatically those that extremize the arc-lenght \cite{shen}.}
\par
One might be tempted at this point to bypass the issue using the Rainbow line-element and find the expression of the four-velocities from $g_{\mu\nu}\dot{q}^\mu\dot{q}^\nu=0$ as in General Relativity.
However we observed earlier that in the Rainbow case the line-element is not invariant under generalized (momentum-dependent) spacetime transformations, therefore this kind of approach does not provide the right particles worldlines.\\ 
A simple example of this feature can be provided using our toy model \eqref{Hamiltonian}, in which, using the Hamiltonian formalism \eqref{HamiltEq} the worldline expression for a massless particle can be easily calculated:
\begin{equation}
x(\eta)-\bar{x}=\int_{\bar{\eta}}^{\eta}\frac{\dot{x}}{\dot{\eta_*}}d\eta_*=\eta-\bar{\eta}-\ell(\beta+\gamma)\Omega\int_{\bar{\eta}}^{\eta} \frac{d\eta_*}{a(\eta_*)}\,.
\end{equation}
The invariance of those worldlines under boost transformations \eqref{BoostTransf} can be easily checked, observing that
\begin{equation}
x(\eta')'-\eta'(1-\ell(\beta+\gamma)\Omega')=x(\eta)-\eta(1-\ell(\beta+\gamma)\Omega)=0\,.
\end{equation}
On the other hand if we try to repeat this procedure with the worldline we obtain from a Rainbow-like light-cone structure $g_{\alpha\beta}dx^\alpha dx^\beta=0$, we find that the result explicitely depends on the rapidity parameter $\xi$
\begin{equation}
x(\eta')'+\eta'\left(1-\ell\frac{\beta+\gamma}{2}\Omega'\right)=-\ell\xi\frac{\beta+\gamma}{2}\Omega\eta\neq 0\,,
\end{equation}
and therefore the Rainbow-metric worldlines are not observer-independent.\\
Again as in the case of the line element \eqref{lineElemInv}, we can recover the right worldlines using the momentum-space metric $\zeta_{\alpha\beta}$ and its light-cone structure $\zeta_{\alpha\beta}dx^\alpha dx^\beta=0$, in fact
\begin{equation}
x(\eta)-\bar{x}=\int_{\bar{\eta}}^{\eta}\sqrt{-\frac{\zeta_{00}}{\zeta_{11}}}d\eta_*=\eta-\bar{\eta}-\ell(\beta+\gamma)\Omega\int_{\bar{\eta}}^{\eta} \frac{d\eta_*}{a(\eta_*)}\,.
\end{equation}
 The reason why this procedure works with momentum-space metric and does not with the Rainbow one is that in this latter case the on-shell relation ${\mathcal H}=0$ does not imply $ds^2=0$. On the other hand, as one can see from the form of the expression of the Hamiltonian in terms of $\zeta$ \eqref{HameLamb}, this does apply for the momentum-space metric line-element which provides the right light-cone structure.

%%%%%%%%%%%%%%%%%%%%%%%%%%%%%%%%%%%%%%%%%%%%%%%%%%%%%%%%%%%%%%%%%%%%%%%%%%%%%%%%%%%%%%%%%%%%%%%%%%%%%%%%%%

\section{Closing remarks}
\label{sec:7}

In this paper we discussed the issues related to the definition of a spacetime metric for theories with Modified Dispersion Relation, with particular attention to the description of the effective spacetime probed by massless particles with energies high enough to test possible Planck-scale effects. A previous approach to this idea was that of Magueijo-Smolin's Rainbow metric \cite{Rainbow}, which has had a large success in Quantum Gravity Phenomenology literature. Here we have shown that in this latter approach the line element is unable to produce Lorentz-deformed-invariant geodesics as the world-lines of the deformed Hamiltonian.

An approach to furnish a coherent picture for four-velocity-dependent spacetime metrics in flat spacetime from the variational point of view, can be found in \cite{Girelli:2006fw} and a study of its DSR realization in \cite{Amelino-Camelia:2014rga}. In these cases the equivalence of the geodesics and the worldlines from Hamilton equations were described, along with its MDR and deformed symmetries, using the language of Finsler geometry. \\
The integration of a few Finsler geometry features in the Rainbow gravity approach could give some guidance on how to overcome some of the limitations that characterize this line of research. For instance we pointed-out in this short paper that generally in the literature the Rainbow geodesic equations are assumed (see {\it exempli gratia} \cite{Rainbow,Ling:2006az,Ali:2014cpa}) to be undeformed, except for the momentum-dependent Christoffel symbols. This assumption is incompatible with the equations obtained from the variation of the action (i.e. the Euler-Lagrange equations) that one gets from a more systematic study. Then, it would be interesting to further investigate on the role that the different connections play in the Deformed Relativistic theories for a massive particles on curved spacetimes, generalizing the analysis presented in \cite{LetiziaLiberatiinprep}.

Despite that approach can be considered as a step forward in the comprehension of  spacetime probed by Planck-scale-sensible particles, the MDR-Finsler metric structure in some cases does not present a well-defined massless limit, which represents a problem for the description of particles with tiny, but in principle finite masses, which could be the case of neutrinos.

Therefore, using a Polyakov-like action for a single particle, we propose a step further in the derivation of this natural geometry, preserving those cited properties of the previous approaches about geodesics and dispersion relations, but with a well-defined massless limit. However we should notice that in the strict sense this is not a Finsler metric, as we lose some properties, like those represented by Eq. (\ref{partial_hom}). A more complete discussion on the spacetime symmetries and the particle's worldlines within this framework can be found in \cite{Loboetalinpreparation2016} for the case of a de Sitter spacetime. We would like to stress that, even though a more careful analysis about the connections for this generalized Finsler metric that we found would be appropriate, we believe is out of the scope of this work and we leave these matters for a future work.

We would like to remark that once the relativity principle is assumed, the Rainbow metrics should be considered as an element of the complex framework described by Relative Locality, in which spacetime is just a mere inference characterized by particle's energies and momenta. In this approach the shape of momentum-space (which is assumed to be curved) influences the particle's dynamics in spacetime, leading to the definition of Planck-scale modified spacetime metrics. 

In this work we intended to set forth the complexity of metric formalism in models with MDR, highlighting how the properties of the metric formalism, which may seem obvious in General Relativity, should not be given for granted in Planck-scale MDR models. Therefore, when approaching Quantum Gravity Phenomenology, one should not just rely on the simple Rainbow metric recipe, but try to balance all the model's ingredients according to the rich theoretical framework here presented, carefully and {\it cum grano salis}. 

%%%%%%%%%%%%%%%%%%%%%%%%%%%%%%%%%%%%%%%%%%%%%%%%%%%%%%%%%%%%%%%%%%%%%%%%%%%%%%%%%%%%%%%%%%%%%%%%%%%%%%%%%%

\section*{Acknowledgements}

{\it The authors would like to thank Giacomo Rosati and Giovanni Amelino-Camelia for useful discussions.} 

I.P.L. is supported by the International Cooperation Program CAPES-ICRANet financed by CAPES - Brazilian Federal Agency for Support and Evaluation of Graduate Education within the Ministry of Education of Brazil grant BEX 14632/13-6. And thanks Conselho Nacional de Desenvolvimento Cienti\'fico e Tecnol\'ogico (CNPq-Brazil) for financial support. 
 
N.L. acknowledges that the research leading to these results has received funding from the European Union Seventh Framework Programme (FP7 2007-2013) under grant agreement 291823 Marie Curie FP7-PEOPLE-2011-COFUND (The new International Fellowship Mobility Programme for Experienced Researchers in Croatia - NEWFELPRO), and also partial support by the H2020 Twinning project n$^{o}$ 692194, RBI-TWINNING.

F.N. acknowledges support from CONACYT grant No. 250298.

\end{document}